\documentclass[aps,onecolumn,prd,showpacs,showkeys,preprintnumbers,superscriptaddress,nobibnotes,notitlepage,floatfix,longbibliography,nofootinbib]{revtex4-2}

\usepackage{graphicx}
\usepackage{amsmath}
\usepackage{amssymb}
\usepackage{xcolor}
\usepackage[colorlinks=true,citecolor=blue,urlcolor=blue]{hyperref}
\usepackage{enumitem}

\def\be{\begin{equation}}
\def\ee{\end{equation}}
\def\bea{\begin{eqnarray}}
\def\eea{\end{eqnarray}}

\def\gev{\, {\rm GeV}}

\newcommand{\gsim}{\lower.7ex\hbox{$\;\stackrel{\textstyle>}{\sim}\;$}}
\newcommand{\lsim}{\lower.7ex\hbox{$\;\stackrel{\textstyle<}{\sim}\;$}}

\newcommand{\cm}{\rm cm}

\newcommand{\km}{\rm km}
\newcommand{\pc}{\rm pc}
\newcommand{\kpc}{\rm kpc}

\newcommand{\s}{\rm s}

\newcommand{\mnras}{MNRAS}
\newcommand{\aap}{A\&A}

\begin{document}

\preprint{CETUP-2024-002, UH511-1336-2024}

\title{ Tools for probing new physics with newly discovered gamma-ray targets
}

\author{Chance Hoskinson}
\affiliation{Department of Physics and Astronomy, University of Utah, Salt Lake City, UT  84112, USA}

\author{Jason Kumar}
\affiliation{Department of Physics and Astronomy, University of Hawai'i, Honolulu, HI 96822, USA}

\author{Pearl Sandick}
\affiliation{Department of Physics and Astronomy, University of Utah, Salt Lake City, UT  84112, USA}

\begin{abstract}
We present a computational tool, \texttt{TweedleDEE}, for empirically modeling diffuse gamma-ray background emission in a $1^\circ$ region of the sky, using publicly available 
gamma-ray data 
off-axis from the region of interest.  This background model allows a user to perform a purely data-driven search for anomalous localized sources 
of gamma-ray emission, including new physics.  
A major application of this tool would be in searching for dark matter annihilation in newly 
discovered astrophysical targets.  For this purpose, we derive a scaling relation for determining velocity-dependent $J$-factors 
using only the stellar parameters, which can be broadly applied to obtain dark matter constraints from new targets.
As an application of these tools, 
we use \texttt{TweedleDEE} and \texttt{MADHATv2} to perform the first search for dark matter annihilation 
in the newly discovered Leo VI dwarf spheroidal galaxy, and present model constraints for a variety of choices of the annihilation channel and velocity dependence of the cross section.
\end{abstract}

\maketitle

\section{Introduction}

One of the most active topics of astrophysics research is the study of localized sources of gamma rays, such as dwarf spheroidal galaxies (dSphs).  
Studies of these objects can inform 
our understanding of known astrophysical objects (e.g.~millisecond pulsars), or could signal the discovery of physics beyond the 
Standard Model (BSM);
dSphs are believed to be dominated by dark matter and are expected to produce few gamma rays through standard astrophysical processes~\cite{Strigari:2012acq}, 
making them  excellent targets for searches for gamma rays produced by BSM processes, such as dark matter particle annihilation~\cite{Aramaki:2022zpw,Boddy:2022knd}.  
New observatories, including the Vera C. Rubin Observatory, are expected to 
discover many new dSphs in the coming years~\cite{LSST:2008ijt,LSSTDarkMatterGroup:2019mwo,Mao:2022fyx}.
In this paper, we will present tools for rapidly analyzing a newly discovered target for BSM sources of gamma-ray emission.

A preliminary step in any study of gamma-ray emission from a localized object is to 
understand diffuse emission in the fore- and background.  We present here \texttt{TweedleDEE}, a Tool for Determining background Emission Empirically, which
generates
a background model for diffuse emission in a $1^\circ$ region of the sky in the $1-100~\gev$ energy range, appropriate for Fermi Large Area Telescope (Fermi-LAT) analyses.
We outline the usage of \texttt{TweedleDEE} in the context of an analysis of a newly discovered compact object as a target for dark matter indirect detection.  As discussed below, the brief study presented here offers a template for rapid analysis of any localized source of gamma rays, both generically and as a dark matter indirect detection target.  

There has been much work on the modeling of diffuse fore- and backgrounds (for simplicity, henceforth referred to collectively as backgrounds) in 
this energy range.  In many studies, the backgrounds are determined by theoretically modeling the processes that produce gamma rays, such as 
high-energy cosmic ray interactions with radiation or gas (see, for example, Refs. ~\cite{Fermi-LAT:2015ycq,Fermi-LAT:2016uux,FermiModel}).   
A alternative approach is to model the background empirically, using gamma-ray data 
off-axis from the location to be studied (see, for 
example, Refs. ~\cite{Geringer-Sameth:2011wse,Geringer-Sameth:2014qqa,Boddy:2018qur}).  This 
empirical approach allows one to separate diffuse emission (which varies slowly across the sky) from localized emission in a way that is agnostic 
to the nature of the diffuse emission processes, and is thus complementary to approaches that model the emission processes themselves.  The tool we present, \texttt{TweedleDEE}, 
follows this empirical approach, and specifically the algorithm of Ref. ~\cite{Geringer-Sameth:2011wse}.

But to translate constraints on anomalous gamma-ray emission in an astrophysical target into 
constraints on dark matter models, one needs an estimate of the dark matter density and velocity 
distributions within 
the target.  The factor encoding this information is known as the 
$J$-factor.  The $J$-factor is typically determined through Jeans modeling, which is computationally 
intensive and requires detailed stellar data.  But especially for the analysis of a newly discovered dSph, 
it is useful to find a simple relation between the $J$-factor and a few coarse-grained stellar observables.
For the case of velocity-independent annihilation, Ref.~\cite{Pace:2018tin} found such a relation between the $J$-factor and three stellar data observables, namely the distance, the half-light radius and the average 
line-of-sight velocity dispersion.  
We generalize this relation to include case of a velocity-dependent annihilation cross section.
   
Even a single new dSph, if large and/or nearby, can dramatically improve the sensitivity of dark matter searches.  
For example, the recent discovery of the objects Carina III~\cite{MagLiteS:2018ylp} and 
Ursa Major III~\cite{2023arXiv231110147S}, 
which may be nearby dSphs, has had a dramatic effect on searches for dark matter 
annihilation.  If either of these objects were confirmed to be a dark matter-dominated dSph, constraints on dark matter annihilation would tighten substantially~\cite{Crnogorcevic:2023ijs,Boddy:2024tiu}.
\texttt{TweedleDEE} produces a background model that can be used directly by the \texttt{MADHATv2} software package \cite{Boddy:2024tiu}, 
which calculates limits on the number of photons emitted by the target object that cannot be attributed to background and is also capable of translating those limits into constraints on any model of new physics that produces emission of photons in the relevant energy range.  Using \texttt{TweedleDEE} and \texttt{MADHATv2}, any newly discovered localized target can 
efficiently be incorporated into a dark matter indirect detection search by a user.

As an illustrative application, we use this tool to develop a diffuse gamma-ray background model for the region of sky encompassing Leo VI, an ultra-faint Milky Way satellite which was discovered very recently~\cite{tan2024pridesatellitesconstellationleo}.  We then use \texttt{MADHATv2}, incorporating this background model, to set limits on the dark matter annihilation cross section. 
Remaining agnostic regarding the properties of an annihilating dark matter candidate, we consider $s$-/$p$-/$d$-wave and Sommerfeld enhanced annihilation (using our determination of the appropriate velocity-dependent $J$-factor), as well as a variety of final states.  

As expected, the dark matter annihilation cross section limits obtained from Leo VI are subleading relative to those obtained 
from previous 
combined analyses, since the estimated $J$-factor of Leo VI is not particularly large.  Moreover, we find that there is no statistically significant 
excess of photons from Leo VI, which is a statement independent of the estimated $J$-factor.  More generally, however, this background model can be incorporated into any 
future combined analysis, and data from dSphs discovered in the future can, in a similar manner, be rapidly incorporated.

The plan of this paper is as follows:  in Sec.~\ref{sec:BgdModel}, we review the algorithm for generating the empirical background model and 
describe the functioning of \texttt{TweedleDEE}.  
In Sec.~\ref{sec:vdep}, we develop a scaling relation that determines the effective $J$-factors for velocity-dependent dark matter annihilation in terms of stellar observables.
In Sec.~\ref{sec:LeoVI}, we apply \texttt{TweedleDEE}
to Leo VI and obtain bounds on 
dark matter annihilation using \texttt{MADHATv2}.  We conclude in Sec.~\ref{sec:conclusion}.

\section{Empirical Background Modeling}
\label{sec:BgdModel}

Here we review the methodology for empirically determining the background model for a region of interest (ROI) 
with $1^\circ$ opening angle.  This methodology largely follows that of Ref. ~\cite{Geringer-Sameth:2011wse} (see also Ref. \cite{Boddy:2019kuw}), 
and is identical to that used in \texttt{MADHATv2} \cite{Boddy:2024tiu}.
We consider photons in the energy range $1 -100\gev$, binned, by default, in 
16 equally spaced logarithmic energy bins.  
We focus on this energy range because it is the range over which the effective area of the Fermi-LAT is roughly constant~\cite{Fermi-LAT:2009ihh}. Furthermore, the energy resolution becomes significantly worse at 
lower energies.  For energies $< 1~\gev$, there is a greater likelihood of a photon being assigned to an 
incorrect energy bin due to energy mismeasurement.  Moreover, the decrease in effective area below $1 \gev$
increases the impact of  energy bin leakage and complicates the calculation of the probability distribution for a signal model.  

An empirically determined background probability distribution can be characterized by a probability mass function (PMF) for an ROI and energy bin.  To construct the PMFs, we generate $10^5$ sample regions of the same size as the ROI, randomly drawn from within $10^\circ$ of the center 
of the ROI.  From these, we discard sample regions which overlap the ROI, which lie within $0.8^\circ$ of an identified 
point source, or which do not lie entirely within $10^\circ$ of the ROI.  The PMF for photon counts in the ROI for each energy bin is given by the normalized histogram of photon 
counts in that energy bin, taken over all of the sample regions described above.
In general, this PMF is not necessarily Poisson.  Furthermore, we have assumed that the PMFs for different energy bins are 
independent; below, in the conclusions, we discuss future work describing deviations from this assumption.

Note that the background models derived in this way will include any emission extended over a 
$\sim 10^\circ$ region.  As such, they are most useful
in searches for emission localized 
to a $\lesssim 1^\circ$ region.  This includes many known dSphs, some of which are relatively 
nearby and have large $J$-factors, such as Ursa Major III~\cite{2024ApJ...961...92S,Crnogorcevic:2023ijs}.  
On the other hand, 
this approach would not provide a suitable background model for emission from Sgr I, for example, which 
spans an angular size of $\sim 10^\circ$ (see, for example, Ref.~\cite{2020MNRAS.497.4162V}).

There are other data-driven approaches to modeling diffuse emission.  A common approach is to 
parametrize the spatial and spectral model of diffuse emission.  A variety of detailed methods can be used to 
determine a background model by maximizing the posterior.  This approach is followed in 
Ref.~\cite{2015A&A...574A..74S}, for example.  Systematic uncertainty in the source model (that is, in the 
expected photon count in a pixel after modeling diffuse sources) can be accounted for by assuming that the mean photon 
count itself is drawn from a Gaussian distribution.  A Fermi-LAT Collaboration study of residuals over the entire Galactic plane found 
that there is a systematic  uncertainty of $\sim 3\%$ in the modeling of diffuse emission, but in any local region this may be larger~\cite{Fermi-LAT:2019yla}.
Our approach differs in that it does not rely on any 
parametrization of the spatial or spectral model for diffuse emission.  Advantages of this approach include its simplicity and computational efficiency.  Moreover, it naturally accounts for any larger deviation between the 
diffuse source model and actual diffuse emission processes in a localized region of the sky.

\subsection{\texttt{TweedleDEE}}

\texttt{TweedleDEE} uses Fermi-LAT photon and spacecraft data to create a Fermi background model in the form of a PMF file and a file containing the photon data from the ROI, both as text files that are also readable by \texttt{MADHATv2}\footnote{MADHATv2 is available at \url{https://github.com/MADHATdm/MADHATv2}} (see  Ref. \cite{Boddy:2024tiu} for details).  The
\texttt{TweedleDEE} package is available at \url{https://github.com/MADHATdm/TweedleDEE}. 
It was tested using \texttt{Python 3.9.19}, and uses the AstroPy, Fermitools, Fermipy, NumPy, and SciPy packages, as specified in the included documentation and environment (\texttt{.yml}) file.   

To construct a background model, \texttt{TweedleDEE} uses 14 years of Fermi-LAT Pass 8 Release 4 data from the mission elapsed time range 239557417 -- 681169985 seconds. Data is selected using \texttt{Fermitools 2.2.0} and \texttt{Fermipy 1.1.6} with the specifications \texttt{evclass=128}, \texttt{evtype=3}, and \texttt{zmax=100}, and the filter \texttt{(DATA\_QUAL>0)\&\&(LAT\_CONFIG==1)}. The instrument response function is taken to be \texttt{P8R3\_SOURCE\_V3}.  Point sources, for masking purposes, are identified using the 4FGL-DR4 source catalog (\texttt{gll\_psc\_v32.fit})~\cite{Fermi-LAT:2022byn,Ballet:2023qzs}, which is automatically downloaded by \texttt{TweedleDEE} on its first run. Exposures for each target region are generated by \texttt{Fermitools}. For each target, Fermi-LAT photon and spacecraft data for the target ROI and its surrounding 10$^\circ$ region must be downloaded by the user via a Fermi-LAT Photon, Event, and Spacecraft Data Query\footnote{\url{https://fermi.gsfc.nasa.gov/cgi-bin/ssc/LAT/LATDataQuery.cgi}}.  Specific query parameters compatible with \texttt{TweedleDEE} along with instructions for downloading the relevant files are available on the \texttt{TweedleDEE} Wiki~\cite{TweedleDEEwiki}.  

After downloading \texttt{TweedleDEE} and the relevant photon and spacecraft files for a target of interest, the user must update the configuration file to set the analysis parameters for \texttt{TweedleDEE}.  A template configuration file is provided with  \texttt{TweedleDEE}.  To analyze a new target, a user can simply update the coordinates of the ROI (matching those used for the photon and spacecraft data query).

\texttt{TweedleDEE} uses \texttt{Fermipy}'s \texttt{GTAnalysis()} function to create  \texttt{ft1\_00.fits} (photon events) and \texttt{ccube\_00.fits} (binned photon counts) files, from which it then extracts the relevant photon event data.   
Within the 10$^\circ$ region around the ROI, $10^5$ random sample regions are chosen by uniform number generation in azimuthal and polar angles.  After pruning sample regions that overlap with the 10$^\circ$ region boundary, the ROI itself, or are located too close on the sky to a known point source, \texttt{TweedleDEE} creates and normalizes a histogram of photon counts across all sample regions for each energy bin.  These normalized histograms are output to a PMF file, in which the first column represents the photon count and each subsequent column is the PMF (normalized histogram) value corresponding to the respective energy bin.  By default, there are 17 columns in a PMF file produced by \texttt{TweedleDEE} for a single target ROI\footnote{Note that it is possible to run \texttt{TweedleDEE} for more than one target simultaneously.  Please refer to \url{https://github.com/MADHATdm/TweedleDEE} for details.}.

\texttt{TweedleDEE} also includes a number of parameters specified in the \texttt{tweedleDEE.py main} function that can be adjusted by a user to suit their analysis.  For example,
\begin{enumerate}[label=(\roman*)]
    \item \texttt{binning = 1}: 1 (default) for 16 logarithmically spaced bins (1-100 GeV); 0 for a single bin
    \item \texttt{Nsample = int(1e5)}: number of random sample regions to generate to construct the PMF
    \item \texttt{target\_size = 10}: radius of the sky region to be analyzed (degrees)
    \item \texttt{sample\_size = 0.5}: radius of the ROI/sample region (degrees)
    \item \texttt{source\_size = 0.8}: radius of the mask for a point source (degrees)
\end{enumerate}

The output of \texttt{TweedleDEE} is a PMF file, which contains the background PMFs for each energy bin, along with a photon event file, which specifies the number of photons detected from the ROI in each energy bin.  This information can be used for a variety of purposes.  In particular, these files are formatted such that they can be used directly by \texttt{MADHATv2} to constrain the number of photons from the ROI that could originate from some source other than the expected background in the sky region.  Furthermore, as described below, if one makes some assumptions about the distribution and microphysics of annihilating dark matter that could have created excess photons, \texttt{MADHATv2} can provide constraints on the dark matter properties.  Below, we present a basic formalism for characterizing the distribution and microphysics of annihilating dark matter in order to obtain dark matter constraints from analyses of localized sources of gamma rays, and use \texttt{TweedleDEE} and \texttt{MADHATv2} to constrain a number of dark matter models.  We note that the relevance of this formalism for characterizing and constraining dark matter extends beyond the limited gamma-ray window utilized here.

\section{ $J$-factors for alternative velocity dependence }
\label{sec:vdep}

In order to constrain dark matter annihilation in a dSph (or other localized target ROI), one must estimate the $J$-factor, 
which is the astrophysical input to the dark matter annihilation rate, and depends on the dark matter density and velocity distributions.  In particular, 
if the dark matter annihilation cross section is given by $\sigma v = (\sigma v)_0 (v/c)^n$, where $v$ is the relative velocity, then 
the differential photon flux arising from dark matter annihilation is given by
\bea
\frac{d\Phi}{dE} &=& \frac{(\sigma v)_0}{8\pi m_X^2} \times J_n \times \frac{dN}{dE} ,
\eea
where $m_X$ is the dark matter particle mass, $dN/dE$ is the photon spectrum per annihilation, and 
\bea
J_n &=& \frac{1}{D^2} \int d^3 r~ d^3 v_1~ d^3 v_2 ~f(\vec{r}, \vec{v}_1) f(\vec{r}, \vec{v}_2) \left(\frac{|\vec{v}_1 - \vec{v}_2|}{c} \right)^n ,
\eea
is the effective $J$-factor.  Here, $D$ is the distance to the dSph, and $f(\vec{r},\vec{v})$ is the velocity distribution of 
dark matter within the subhalo ($\int d^3v~f(\vec{r}, \vec{v}) = \rho (\vec{r})$).  We have assumed that the size of the subhalo 
is much smaller than $D$, and that the dark matter particle 
is its own antiparticle.
The effective $J$-factor includes all dependence of the flux on the astrophysics of the target.  An estimate of the $J$-factor for Leo VI in the case of annihilation from an $s$-wave initial state ($n=0$) is presented in 
Ref.~\cite{tan2024pridesatellitesconstellationleo}.

Since the velocity dependence of any dark matter annihilation is currently unknown, we also consider dark matter annihilation under the assumption of a velocity-dependent 
dark matter annihilation cross section.  This is particularly interesting for dSphs that have a relatively small velocity dispersion, as is the case for Leo VI. 
As a result, its $J$-factor can be 
enhanced, relative to other targets, in scenarios in which the annihilation cross section 
grows with decreasing relative velocity.  To study these cases, we must generalize the $s$-wave $J$-factor 
to the velocity-dependent case.

We assume that the dark matter density profile is of the Navarro-Frenk-White (NFW) form, characterized by 
\bea
\rho (r) &=& \frac{\rho_s}{ (r/r_s) (1 + r/r_s)^2} ,
\eea
where $r_s$ is the scale radius, and $\rho_s$ is the scale density.  We replace $\rho_s$ with the velocity parameter 
$v_0 = \sqrt{4\pi G_N \rho_s r_s^2}$, where $G_N$ is Newton's constant.

Here we  consider four cases for the scaling of the dark matter annihilation cross section: $\sigma v = (\sigma v)_0~ (v/c)^n$,  
where $n=-1, 0, 2$ or $4$.
Note that $n=0$ is the standard case in which dark matter annihilates from an $s$-wave 
initial state, and $J_0$ corresponds to the usual $J$-factor.  
The $n=2$ or $n=4$ cases arise if dark matter instead annihilates dominantly from a $p$- or $d$-wave 
initial state, respectively.  The $n=-1$ case arises if dark matter annihilation is Sommerfeld enhanced as a result of 
a long-range attractive force  between the dark matter particles~\cite{Arkani-Hamed:2008hhe,Feng:2010zp}.  It is this last case which is generally the most interesting 
for indirect detection searches of dark matter subhalos, since subhalos tend to have a relatively small velocity dispersion, 
leading to a larger Sommerfeld enhancement of the annihilation rate.

For a target such as Leo VI, which is small compared to its distance from the solar system,
its $J$-factor integrated over a $1^\circ$ aperture is 
well-approximated by its total $J$-factor.
The $J$-factors can essentially be determined from dimensional analysis, and are given by~\cite{Boddy:2017vpe,Boddy:2019wfg}
\bea
J_n 
&=& \frac{1}{4\pi G_N^2} \frac{v_0^4}{D^2 r_s} (v_0/c)^n ~ \tilde J_n ,
\nonumber\\
 &=& \left(1.2 \times 10^{16} \gev^2 \cm^{-5} \right) \left(\frac{v_0}{5~\km~\s^{-1}} \right)^4 
\left(\frac{D}{100~\kpc} \right)^{-2}
\left(\frac{r_s} {100~\pc} \right)^{-1} (v_0/c)^n ~ \tilde J_n .
\label{eq:Jn_halo}
\eea
The constants $\tilde J_n$ depend on the form of the density profile, and for an NFW profile 
(assuming a velocity-distribution which depends only on energy) 
are given by~\cite{Boucher:2021mii}
\bea
\tilde J_{-1} &=& 0.83 ,
\nonumber\\
\tilde J_{0} &=& 1/3 ,
\nonumber\\
\tilde J_{2} &=& 0.14 ,
\nonumber\\
\tilde J_{4} &=& 0.12 .
\eea

Unfortunately, Eq.~\ref{eq:Jn_halo} depends on parameters of the dark matter halo, namely $r_s$ and $v_0$, which are difficult to determine from 
observations.  The halo parameters (and, in turn, the gravitational potential) are generally fit by Jeans modeling of the observed stellar kinematics, which is a computationally intensive process for which detailed data may not be immediately available.   
To make contact directly with stellar kinematic parameters, we use a scaling relation for the $s$-wave $J$-factor presented in 
Ref.~\cite{Pace:2018tin}.
In particular, Ref.~\cite{Pace:2018tin} found that $J$-factors for a sample of dSphs obtained by fitting the gravitational potential could in turn 
be expressed in terms of stellar kinematic parameters as 
\bea
J_0 &=& \left(10^{17.87}  \gev^2 \cm^{-5} \right) \left(\frac{\sigma_{los}}{5~\km~\s^{-1}} \right)^4 
\left(\frac{D}{100~\kpc} \right)^{-2}
\left(\frac{r_{1/2}}{100~\pc} \right)^{-1} , 
\label{eqn:SwaveScalingRelation}
\eea
where $\sigma_{los}$ is the line-of-sight velocity dispersion and $r_{1/2}$ is the half-light radius.  Since the only relevant halo 
parameters are $v_0$ and $r_s$, dimensional analysis implies that $\sigma_{los} \propto v_0$ and $r_{1/2} \propto r_s$.

The expression in Eq.~\ref{eqn:SwaveScalingRelation} matches an analytic derivation \cite{Pace:2018tin} 
(using the half-mass estimator \cite{Walker:2009zp,Wolf:2009tu})
for $r_{1/2} = 0.25~r_{s}$, 
which would imply $v_0 \sim 5.2~\sigma_{los}$.  Of course, this relationship between stellar parameters and halo parameters 
is not exact, and the $J$-factors determined through Jeans modeling exhibit scatter about the relation given in Eq.~\ref{eqn:SwaveScalingRelation}.  
Indeed, using the central values of the stellar kinematic parameters for Leo VI identified in Ref.~\cite{tan2024pridesatellitesconstellationleo}, namely 
$\sigma_{los} = 2.85~ \km / \s$, $r_{1/2} = 90~\pc$, and taking $D = 111~\kpc$, one finds from 
Eq.~\ref{eqn:SwaveScalingRelation} that $J_0 = 10^{16.85} \gev^2 \cm^{-5}$, which is well within $1\sigma$ of the 
value obtained in Ref.~\cite{tan2024pridesatellitesconstellationleo} directly through Jeans modeling 
($\log_{10} (J_0/\gev^2 \cm^{-5}) = 17.1^{+0.9}_{-1.2}$)\footnote{This $J$-factor is obtained by 
integrating  over $0.5^\circ$ (the size of our ROI).  Note however that the $J$-factor integrated 
over $1^\circ$ is the same, as this angular scale essentially encompasses the entire dSph.}.
For our purposes, this scaling relation will be sufficient.

Using the relation between halo parameters and stellar kinematic parameters,  we arrive at a scaling relation applicable 
for any choice of power-law velocity-dependence of the cross section:
\bea
J_n &=& \left(10^{17.87}  \gev^2 \cm^{-5} \right) \left(\frac{\sigma_{los}}{5~\km~\s^{-1}} \right)^4 
\left(\frac{D}{100~\kpc} \right)^{-2}
\left(\frac{r_{1/2}}{100~\pc} \right)^{-1}  \left(\frac{5.2~\sigma_{los}}{c} \right)^n \frac{\tilde J_n}{\tilde J_0}.
\label{eqn:Jn}
\eea
Note that, although we focus in this work on Leo VI, Eq.~\ref{eqn:Jn} can be used to determine 
the velocity-dependent effective $J$-factor for any dSph whenever the distance to the dSph, half-light radius 
and line-of-sight velocity dispersion are available.
Although the intrinsic scatter in this relation is relatively small~\cite{Pace:2018tin}, 
the uncertainty in the input parameters is significant.  We will 
therefore adopt an uncertainty of plus or minus an order of magnitude, which is approximately the uncertainty reported 
by Ref.~\cite{tan2024pridesatellitesconstellationleo} for $J_0$.

The scaling relation above gives us
\bea
\log_{10} (J_{-1} / \gev^2 \cm^{-5} ) &=& 21.55^{+0.97}_{-0.81} , 
\nonumber\\
\log_{10} (J_{2} / \gev^2 \cm^{-5} ) &=& 7.86^{+0.97}_{-0.81} , 
\nonumber\\
\log_{10} (J_{4} / \gev^2 \cm^{-5} ) &=& -0.82^{+0.97}_{-0.81} .
\eea
Note that, although the $s$-wave $J$-factor for Leo VI is substantially smaller than those for almost all of the 25 dSphs 
considered in Ref.~\cite{Boddy:2019qak}, this is less so for the Sommerfeld enhanced $J$-factor.

We also apply this formalism to another recently discovered dSph, Ursa Major III~\cite{2024ApJ...961...92S}.  
Because it is so nearby ($D = 10 \pm 1~\kpc$), it has a large $J$-factor (subject, of course, to uncertainties in its dark matter content), and can produce much more sensitive constraints on dark matter 
annihilation than Leo VI. 
Constraints on $s$-wave dark matter annihilation in Ursa Major III have been obtained~\cite{Crnogorcevic:2023ijs,Boddy:2024tiu}, and more 
recently, the $J$-factors associated with Sommerfeld enhanced and $p$-wave annihilation were obtained through Jeans modeling~\cite{Zhao:2024say}.  
We can thus compare our simple scaling result to the result obtained through Jeans modeling.
Using 11 member 
stars identified in Ref.~\cite{2024ApJ...961...92S} (including two stars which may be velocity outliers), one finds $r_{1/2} = 3 \pm 1~\pc$  
and $\sigma_{los} = 3.7^{+1.4}_{-1.0}~\km / \s$. 
Using Eq.~\ref{eqn:Jn}, we find  $\log_{10} (J_{-1} / \gev^2 \cm^{-5}) = 25.6^{+0.7}_{-0.5}$, where the uncertainties are obtained by 
propagating the uncertainties in the parameters.  
This lies within the error bars of the result presented in Ref.~\cite{Zhao:2024say} 
$\left( \log_{10} (J_{-1} / \gev^2 \cm^{-5}) = 25.3^{+0.5}_{-0.5}\right)$.

It thus appears that the scaling relation in Eq.~\ref{eqn:Jn} can be used to provide a robust estimate for the velocity-dependent 
$J$-factor, given only two stellar parameters ($r_{1/2}$ and $\sigma_{los}$), without requiring a full Jeans analysis.

\section{An Application: Constraints on Dark Matter Annihilation in Leo VI 
}
\label{sec:LeoVI}

Here, we use \texttt{TweedleDEE} and \texttt{MADHATv2} to constrain dark matter annihilation in the recently discovered dSph Leo VI~\cite{tan2024pridesatellitesconstellationleo}.   We use the $J$-factor formalism above to analyze four scenarios for velocity dependence of the annihilation cross section, and choose four annihilation final states, which correspond to four benchmark gamma-ray spectra.

\begin{figure}[h]
    \centering
    \includegraphics[width=0.5\textwidth]{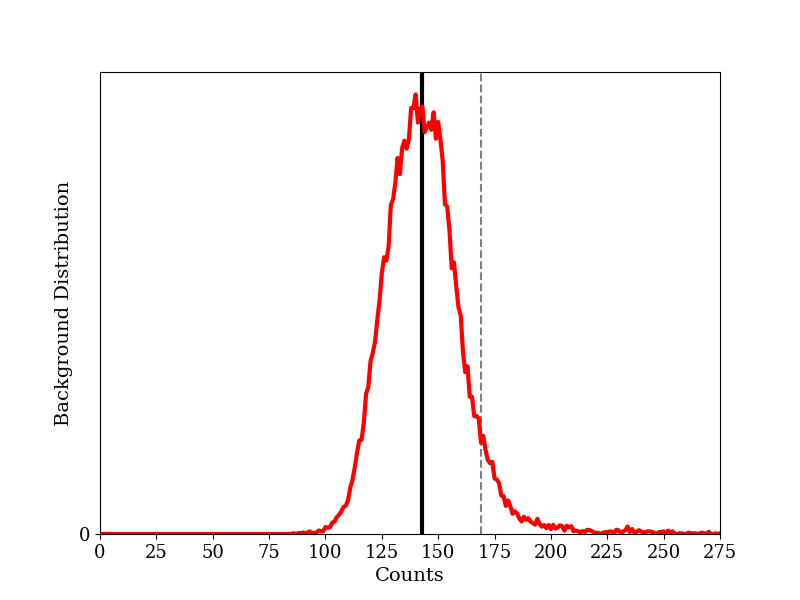}
    \caption{Background distribution for photons in the energy range $1-100~\gev$ for Leo VI generated with \texttt{TweedleDEE}.  The vertical black line at 143 represents the number of observed photons in the $1-100$ GeV energy range from the ROI.
    The light gray dashed line is at 169, and the normalized area under the red curve to the left of the light gray line is $0.95$.
   }
    \label{fig:pmf}
\end{figure}

The PMF for Leo VI generated with \texttt{TweedleDEE} for the energy range $1-100~\gev$ is displayed in Figure~\ref{fig:pmf} as a red histogram.  
The total number of photons observed from the ROI in the $1-100~\gev$ energy range 
during this time period is 143, represented by the vertical black line.  From the background model (treating the $1 - 100~\gev$ energy 
range as a single bin), the number of photons attributable 
to diffuse backgrounds is $\leq 169$, at $95\%$ confidence level (CL).  As such, there is no statistically significant 
excess of photons from the direction of Leo VI beyond what may be attributable to diffuse backgrounds.  
Note that this statement is entirely data-driven, and is independent of any estimate of the $J$-factor.

\begin{figure}[tb]
    \centering
    \includegraphics[width=0.5\textwidth]{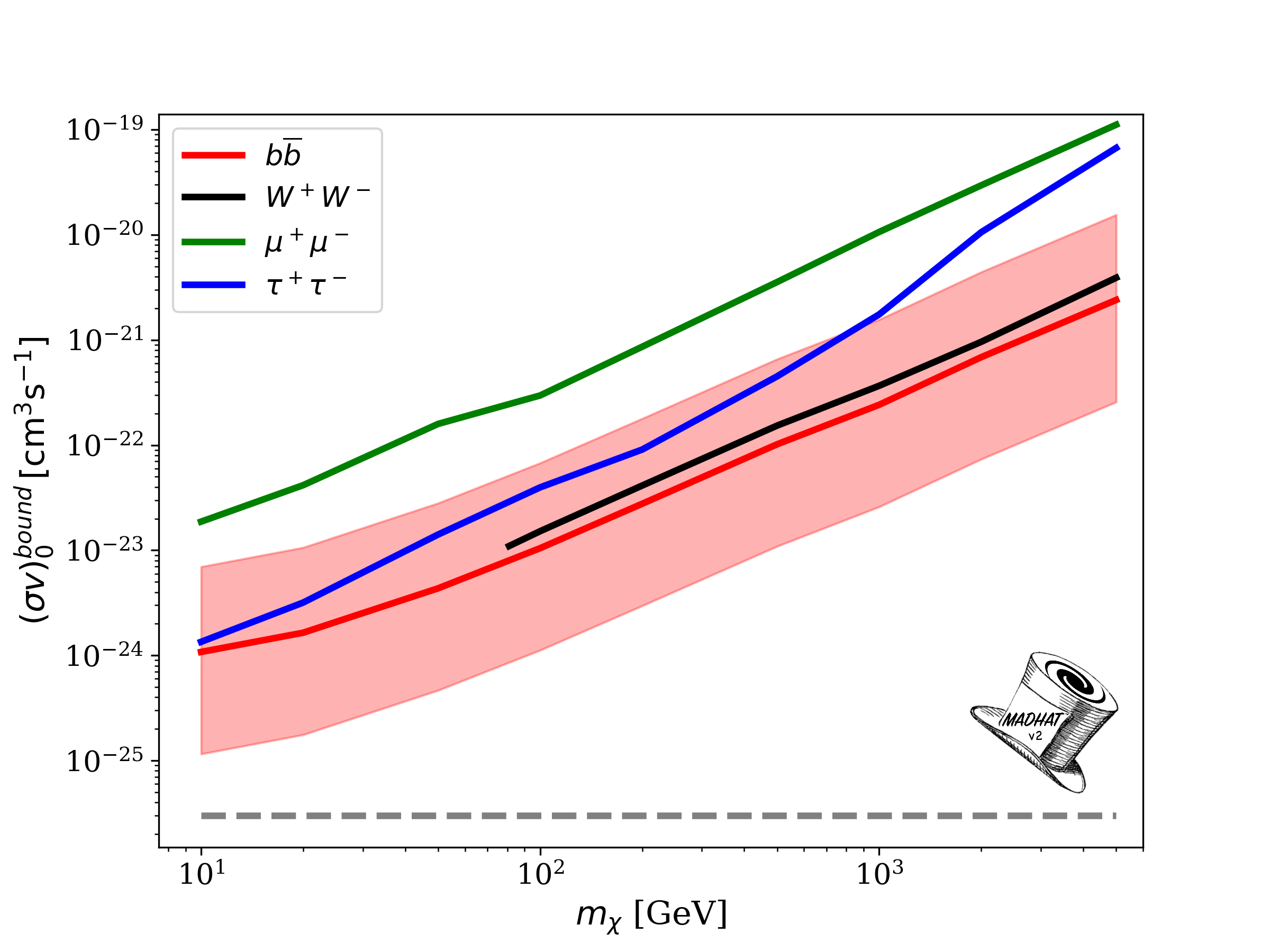}
    \caption{95\% CL $(\sigma v)_0$ for $s$-wave annihilation in Leo VI to the final states $b\overline{b}$ (red), $W^+W^-$ (black), $\mu^+\mu^-$ (green), and $\tau^+\tau^-$ (blue) using \texttt{MADHATv2}.
    The thermal cross section is included as a gray, dashed horizontal line.
     The red band indicates the variation of the bounds in the $b \bar b$ channel due to systematic uncertainty 
in the $J$-factor. 
   }
    \label{fig:four_two_body_plot}
\end{figure}

We obtain constraints on dark matter annihilation in Leo VI using \texttt{MADHATv2}, with the background model 
obtained from \texttt{TweedleDEE} using 16 equal logarithmically spaced energy bins.
We assume that the dark matter particle is its own antiparticle and consider four two-body final states. Photon spectra for these final states is obtained from Refs.~\cite{Ciafaloni:2010ti,Cirelli:2010xx}.  
In Figure~\ref{fig:four_two_body_plot} we present $95\%$ CL bounds in the $(m_X, (\sigma v)_0)$ plane assuming 
$s$-wave annihilation to the final states $b \bar b$ (red), $W^+ W^-$ (black), $\mu^+ \mu^-$  (green) 
and $\tau^+ \tau^-$ (blue).  The red band indicates the variation of the bounds in the $b \bar b$ channel due to systematic uncertainty 
in the $J$-factor.  
For reference, we also present the typical thermal annihilation cross section of $\sigma v = 3\times 10^{26}$ cm$^3$s$^{-1}$ as a gray dashed line.

\begin{figure}[h]
    \centering
    \includegraphics[width=0.48\textwidth]{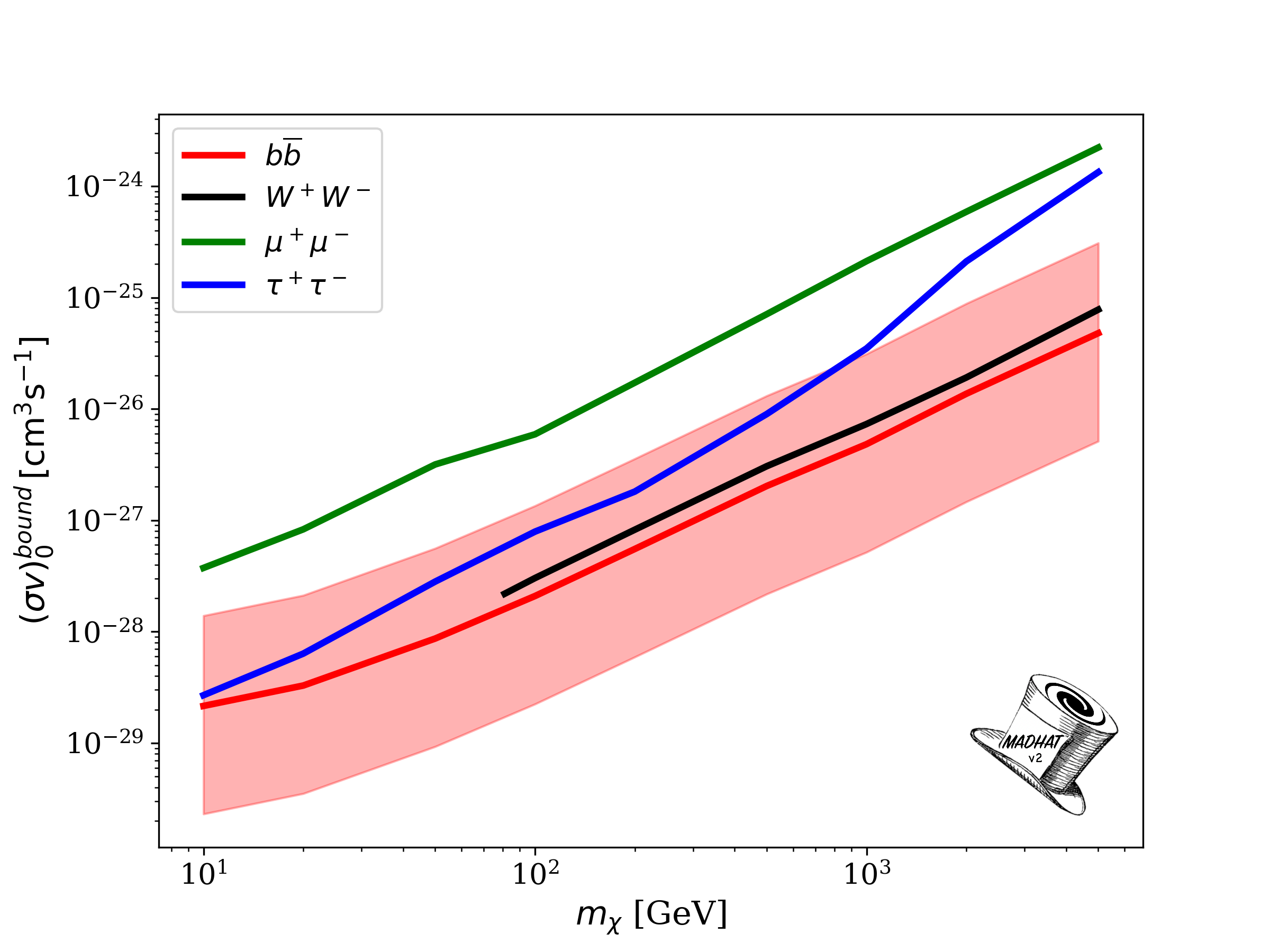}
    \includegraphics[width=0.48\textwidth]{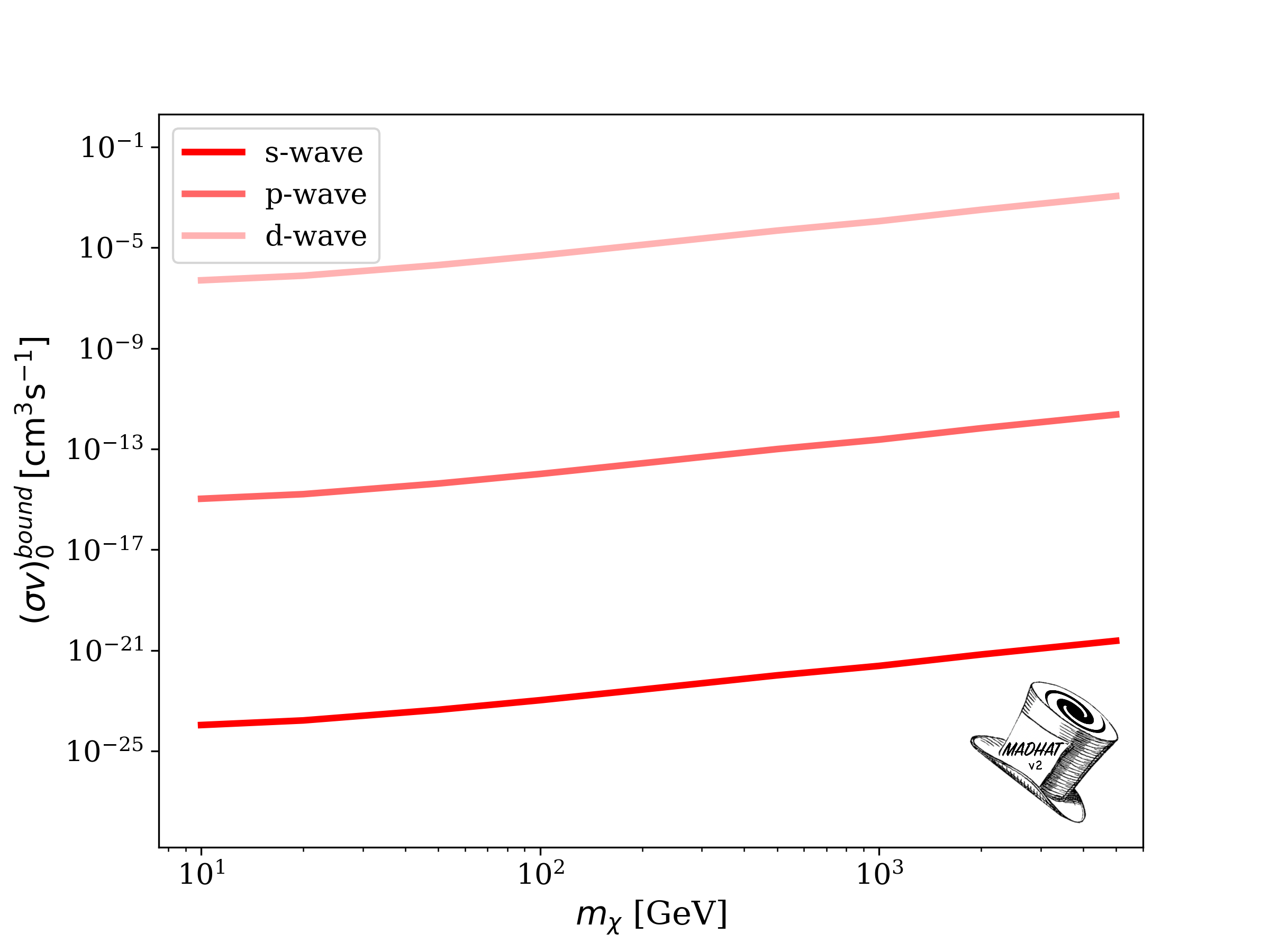}
    \caption{95\%-confidence bounds on $(\sigma v)_0$ for Sommerfeld enhanced dark matter annihilation (left panel), and for $p$-wave and $d$-wave annihilation (right panel) in Leo VI using \texttt{MADHATv2}. In the left panel, constraints are presented for  
  annihilation to the final states $b\overline{b}$ (red), $W^+W^-$ (black), $\mu^+\mu^-$ (green), and $\tau^+\tau^-$ (blue).  The red band indicates the variation of the bounds in the $b \bar b$ channel due to systematic uncertainty in the $J$-factor. 
   In the right panel, we show the constraints on the annihilation cross section to $b \bar b$ for $s$-wave (red, as in Fig.~\ref{fig:four_two_body_plot}), $p$-wave (lighter red), and $d$-wave (lightest red) annihilations.
   The red band in the left panel indicates the variation of the bounds in the $b \bar b$ channel due to systematic uncertainty in the $J$-factor. 
   }
    \label{fig:vdep}
\end{figure}

In Figure~\ref{fig:vdep}, we present constraints on the dark matter annihilation cross section assuming Sommerfeld enhanced dark matter annihilation (left panel), and for
$p$-wave and $d$-wave annihilation (right panel).  In the left panel of Fig.~\ref{fig:vdep}, we show the constraints on the annihilation cross section assuming Sommerfeld enhanced dark matter annihilation to the final states $b \bar b$ (red), $W^+ W^-$ (black), $\mu^+ \mu^-$  (green) and $\tau^+ \tau^-$ (blue).  The red band indicates the variation of the bounds in the $b \bar b$ channel due to systematic uncertainty in the $J$-factor.  
In the right panel of Fig.~\ref{fig:vdep}, we show the constraints on the annihilation cross section to $b \bar b$ for $s$-wave (red, as in Fig.~\ref{fig:four_two_body_plot}), $p$-wave (lighter red), and $d$-wave (lightest red) annihilations.

These constraints represent the first constraints on dark matter annihilation from Leo VI.  While these constraints are not the strongest constraints on the dark matter parameter space explored, they serve as an application of the formalism presented here for analyses of dark matter annihilation in new targets.  This formalism could also be used to incorporate Leo VI and/or other newly discovered targets into a combined analysis of many targets, which could result in more stringent bounds and may mitigate some of the $J$-factor uncertainties\footnote{We note that, although $J$-factor uncertainties could be mitigated in a combined analysis, this is not necessarily the 
case, for example, if the $J$-factors were systematically over- or underestimated.}.

\section{Conclusion}
\label{sec:conclusion}

We have developed and made publicly available \texttt{TweedleDEE}, a tool for empirically generating a background model for 
diffuse gamma-ray background emission, for use in searches for a localized gamma-ray source.  \texttt{TweedleDEE} 
generates an energy-dependent background model in the $1 - 100~\gev$ energy range using Fermi-LAT data taken 
off-axis from the ROI.  
This entirely data-driven background model is  agnostic to the physical processes responsible for diffuse gamma-ray emission, 
and thus complements background models derived from theoretical modeling of  emission processes.

An obvious application for this type of background model is in the indirect detection of dark matter annihilating 
in dwarf spheroidal galaxies.
This tool produces a background model in a form which can be directly incorporated into the \texttt{MADHATv2} software package.

To constrain dark matter annihilation for generic cases in which the annihilation cross section may exhibit velocity-dependence, one 
must determine the appropriate effective $J$-factor.  Assuming an NFW profile, we have used dimensional analysis to generalize 
the results of Ref. \cite{Pace:2018tin}, obtaining a scaling relation for the effective $J$-factor in terms of the 
distance, the half-light radius 
and the line-of-sight velocity dispersion.  This expression is of general utility, and allows one to estimate the effective $J$-factor 
for a variety of choices of the cross section velocity-dependence, using only a few stellar observables and without full Jeans modeling.

Here we have used \texttt{TweedleDEE} and \texttt{MADHATv2} to obtain the first constraints on dark matter annihilation in 
the newly discovered dSph Leo VI, for the cases of $s$-wave, $p$-wave, $d$-wave, and Sommerfeld enhanced annihilation, and for four benchmark two-body final states.  We find, in particular, that there is no statistically significant excess of photons 
from Leo VI, beyond what may be attributed to diffuse backgrounds, and obtain constraints on the dark matter model parameter space.

There are some interesting topics for future work.  Although a disadvantage of this type of background model is that it does 
not take advantage of our knowledge of processes which can produce diffuse gamma rays, an advantage is that it may be more 
robust to mismodeling of those processes.  It would be interesting to perform a detailed comparison of theoretically motivated 
and purely empirical background models, to see which model provides a better fit to Fermi data.

We have made the simplifying assumption that the photon count distributions in different energy bins are independent.  In an 
empirical background model, this generally is not true~\cite{Geringer-Sameth:2014qqa,Boddy:2024tiu}.  A source of photons in a sampling region may generate photons in multiple energy bins, resulting in a correlation of the resulting photon count distributions.  The authors of this paper are currently considering this scenario in detail.

\section*{Acknowledgments}

We express our gratitude and appreciation to Kimberly K.~Boddy for collaboration on Ref.~\cite{Boddy:2018qur}, for which she originally developed code that implements the sky selections using Fermitools, and which formed part of the basis for \texttt{TweedleDEE}.  
We are also grateful to Zachary Carter, Louie Strigari and Natalia Tapia-Arellano for useful discussions.  
JK is supported in part by DOE Grant No. DE-SC0010504.
PS is supported in part by NSF Grant No. PHY-2412834.  
JK and PS wish to acknowledge the Center for Theoretical Underground Physics and Related Areas (CETUP*), the Institute for Underground Science at Sanford Underground Research Facility (SURF), and the South Dakota Science and Technology Authority for hospitality and financial support, as well as for providing a stimulating environment where part of this work was initially conceived.

\bibliography{LeoVI}

\end{document}